\documentclass{aastex631}

\usepackage{amssymb}
\usepackage{amsmath}
\usepackage{gensymb}
\usepackage{graphicx}
\usepackage{xcolor}
\usepackage{natbib}

\newcommand{\hi}{H{\sc i}}

\newcommand{\mh}{\rm M_{H{\textsc i}}}
\newcommand{\MHI}{\rm M_{H{\textsc i}}}
\newcommand{\Mstar}{\rm M_*}
\newcommand{\mb}{\rm M_B}
\newcommand{\hii}{H{\sc i} 21\,cm}
\newcommand{\msun}{\rm M_{\odot}}

\newcommand{\kms}{km~s$^{-1}$}

\shorttitle{H{\sc i} scaling relations at $z \approx 0.35$}
\shortauthors{A. Bera et al.}
\graphicspath{{./}{figures/}}

\begin{document}

\title{\Large{Atomic hydrogen scaling relations at $z \approx 0.35$}}

\correspondingauthor{Nissim Kanekar}
\email{nkanekar@ncra.tifr.res.in}

\author{Apurba Bera}
\affil{International Centre for Radio Astronomy Research, Curtin University, Bentley, WA 6102, Australia}
\affil{National Centre for Radio Astrophysics, Tata Institute of Fundamental Research, Pune 411007, India}
\affil{Inter-University Centre for Astronomy and Astrophysics, Pune 411007, India}

\author{Nissim Kanekar}
\affil{National Centre for Radio Astrophysics, Tata Institute of Fundamental Research, Pune 411007, India}

\author{Jayaram N. Chengalur}
\affil{National Centre for Radio Astrophysics, Tata Institute of Fundamental Research, Pune 411007, India}

\author{Jasjeet S. Bagla}
\affil{Indian Institute of Science Education and Research Mohali, Knowledge City, Sector 81, Sahibzada Ajit Singh Nagar, Punjab 140306, India}


\begin{abstract}

The atomic hydrogen (H{\sc i}) properties of star-forming galaxies in the local Universe are known to correlate with other galaxy properties via the ``H{\sc i} scaling relations''. The redshift evolution of these relations  serves as an important constraint on models of galaxy evolution. However, until recently, there were no estimates of the H{\sc i} scaling relations at cosmological distances. Using data from a deep Giant Metrewave Radio Telescope H{\sc i} 21\,cm survey of the Extended Groth Strip, and the technique of spectral line stacking, we determine the scaling relation between the H{\sc i} mass and the stellar mass for star-forming galaxies at $z\approx0.35$. We use this measurement, along with the main-sequence relation in galaxies, to infer the dependence of the \hi\ depletion timescale of these galaxies on their stellar mass. We find that massive star-forming galaxies at $z\approx0.35$, with stellar mass $\rm M_* \gtrsim10^{9.5}\:M_{\odot}$, are H{\sc i}-poor compared to local star-forming galaxies of a similar stellar mass. However, their characteristic H{\sc i} depletion time is lower by a factor of $\approx 5$ than that of their local analogues, indicating a higher star-formation efficiency at intermediate redshifts (similar to that at $z \approx 1$).  While our results are based on a relatively small cosmic volume and could thus be affected by cosmic variance, the short characteristic \hi\ depletion timescales ($\lesssim 3$~Gyr) of massive star-forming galaxies at $z \approx 0.35$  indicate that they must have acquired a significant amount of neutral gas through accretion from the circumgalactic medium over the past four Gyr, to avoid quenching of their star-formation activity.

\end{abstract}


\section{Introduction} \label{sec:intro}

The atomic hydrogen (\hi) reservoir of galaxies provides the primary fuel reservoir for star-formation activity, and is thus a critical factor in galaxy evolution. While star-formation takes place in the molecular phase and the exact role of \hi\ in galaxy evolution is not clearly understood, the lack of \hi\ in red quenched galaxies in the local Universe suggests that the exhaustion of the \hi\ reservoir leads to the suppression, and subsequent quenching, of the star-formation activity in galaxies \citep[e.g.][]{saintonge22araa}. Conversely, \hi\ has been detected in a significant fraction ($\approx 40$\%) of early-type galaxies in the Atlas3D sample \citep[e.g.][]{serra12mnras}, indicating that the conversion of \hi\ to stars is likely to be affected by several factors \citep[e.g.][]{davis14mnras}. Measurements of the \hi\ content of galaxies via \hii\ spectroscopy have hence long been of much interest in probing galaxy evolution.

In the nearby Universe, the \hi\ properties of star-forming galaxies are known to correlate with their stellar properties\footnote{Note that the correlations typically have significantly higher scatter for volume-limited samples that contain both star-forming and passive galaxies \citep[see, e.g.,][]{parkash18apj}.}, through the \hi\ scaling relations \citep[e.g.][]{toribio11apj2,denes14mnras, wang16mnras,Romeo2020}. These include relations between the \hi\ mass, $\MHI$, and the stellar mass, $\Mstar$ \citep[e.g.][]{catinella18mnras, parkash18apj}, the \hi\ mass and different optical magnitudes \citep[e.g.][]{denes14mnras}, the \hi\ mass and the \hi\ size \citep[e.g.][]{broeils97aa,wang16mnras}, etc. Such relations provide a critical constraint on numerical and semi-analytical models of galaxy evolution \citep[e.g.][]{dave2020}. 

Unfortunately, the intrinsic faintness of the \hii\ transition has meant that there have been few detections of \hii\ emission from individual galaxies at $z \gtrsim 0.25$ \citep[e.g.][]{catinella15mnras,fernandez16apj,gogate20mnras}. Further, even these detections have only been in the most massive galaxies; this has meant that we have had no information about \hi\ scaling relations beyond the local Universe. Recently, the \hii\ stacking technique has provided a unique way to statistically measure the average \hi\ properties of a population of galaxies \citep[e.g.][]{zwaan00thesis, chengalur01aa, jaffe16mnras, bera19apjl,chowdhury21apjl,chowdhury22survey}. This has made it possible to measure the \hi\ scaling relations for cosmologically-distant galaxies, at both intermediate \citep[$z \approx 0.4$;][]{bera22apjl, sinigaglia22apjl} and moderately-high \citep[$z \approx 1$;][]{chowdhury22scaling} redshifts. In this {\it Letter}, we report a measurement of the $\MHI - \Mstar$ scaling relation in star-forming galaxies at $z \approx 0.35$, based on deep Giant Metreware Radio Telescope (GMRT) \hii\ spectroscopy  of the Extended Groth Strip (EGS). Comparing our measurement to the known scaling relation in the local Universe, we quantify the evolution of the \hi\ reservoir and the \hi\ depletion timescale (and thus, the star-formation efficiency) of galaxies over the past four Gyr.\footnote{Throughout this work, we use a flat $\Lambda$-cold dark matter ($\Lambda$CDM) cosmology, with ($\rm H_0$, $\rm \Omega_{m}$, $\rm \Omega_{\Lambda})=(70$~km~s$^{-1}$~Mpc$^{-1}$, $0.3, 0.7)$.  All magnitudes in this work are in the AB system \citep{oke74apjs}.}

\section{Observations and data processing} \label{sec:data}

We used the GMRT Band-5 receivers to carry out a $\approx 350$-hr observation of the EGS between March 2017 and June 2019, in proposals 31\_038 (P.I.: J.~S.~Bagla), 34\_083 (P.I.: N.~Kanekar), 35\_085 (P.I.: A.~Bera), and 36\_064 (P.I.: A.~Bera). The observations and data analysis are described in detail by \citet{berainprep}; a brief summary is provided below. 

We used the GMRT Wideband Backend as the correlator for our observations of the EGS, with a bandwidth of 400~MHz covering the frequency range $970 - 1370$~MHz, and sub-divided into 8,192 spectral channels. The initial data editing, gain calibration, and bandpass calibration were carried out in the classic {\sc aips} package \citep{greisen03book}, following standard procedures. Imaging and self-calibration were carried out independently for each observing cycle, and the inferred antenna-based gains applied to the spectral-line visibilities. For each cycle, the model visibilities corresponding to the continuum image of the cycle were then subtracted out from the calibrated spectral-line visibilities. The residual visibilities were then imaged, again independently for each cycle, using the task {\sc tclean} in the {\sc casa} package \citep[version 5.6;][]{mcmullin07book} to make the final spectral cubes. We emphasize that independent spectral cubes were made for each observing cycle; this was done to ensure that any low-level radio frequency interference (RFI) in a given cycle would not affect the data from the other cycles. The cubes were made in the barycentric frame, using w-projection \citep{cornwell08} and Briggs weighting with {\sc robust}=0.5 \citep{briggs95}, and were corrected for the frequency-dependent shape of the GMRT primary beam. The FWHM of the GMRT primary beam is $\approx 28' - 34'$ over the frequency range $1184-1000$~MHz, corresponding to a spatial scale of $\approx 5.5 - 11.3$~Mpc over the redshift range $z=0.20-0.42$. The cubes have a frequency resolution of 97.7~kHz, equivalent to a velocity resolution of $\approx 21-30$~\kms\ across the observing band. The native angular resolutions of the cubes (i.e. the FWHMs of the synthesized beams) are $\approx 2\farcs6 - 3\farcs3$, corresponding to spatial resolutions of $\approx 9 - 18$~kpc for the redshift range $z = 0.20-0.42$.

\section{\hii\ stacking analysis} \label{sec:stacking}

\subsection{The galaxy sample} \label{subsec:sample}

\hii\  stacking experiments critically require a large sample of galaxies within the telescope primary beam, with accurately measured positions and redshifts \citep[e.g.][]{chowdhury22survey}. The redshift accuracy needs to be significantly better than the typical \hii\ line width in order to properly align the line emission from the different galaxies; this implies a required redshift accuracy of $\lesssim 100$~\kms\ \citep[e.g.][]{maddox13}. Our target field, the EGS, has excellent spectroscopic coverage from the DEEP2 and DEEP3 surveys \citep{newman13apjs, cooper12mnras}, which provide spectroscopic redshifts (accurate to $\lesssim$ 62~\kms, for redshift quality code $\rm Q \geq 3$) for a large number of galaxies with $\rm R_{AB} \leq 24.1$. The DEEP2 and DEEP3 spectroscopic redshift catalogues are $\approx$ 60\% complete at the limiting apparent magnitude $\rm R_{AB} = 24.1$ for $\rm Q \geq 3$ \citep[see Fig.~32 of][]{newman13apjs}. Most of the incompleteness arises from galaxies with faint emission lines, which are likely to lie at $z > 1.4$ \citep{newman13apjs}. However, we note that it is possible that low-redshift galaxies with faint optical emission lines (e.g. obscured, dusty galaxies) might also be missing from our sample.

The parent sample for our \hii\ stacking experiment was selected from the combined catalogue of the DEEP2 and the DEEP3 surveys\footnote{http://deep.ps.uci.edu/deep3}. The redshift range of our sample was restricted to $0.20 \leq z \leq 0.42$; the upper limit is determined by the frequency coverage of the GMRT Band-5 receivers, and the lower limit by the DEEP2 and DEEP3 redshift coverage. The sample was further restricted to the 808 galaxies which (1)~lie within the FWHM of the uGMRT primary beam at the redshifted \hii\ line frequency\footnote{This was done because systematic effects in the spectral data cubes are often significant outside the FWHM of the primary beam.}, (2)~have reliable redshift estimates \citep[quality code $\geq 3$;][]{newman13apjs}, and (3)~have absolute B-band magnitudes of $\mb\ \leq -16$.\footnote{This $\mb\ $ limit was chosen to exclude very faint galaxies from our sample. These galaxies are expected to have low \hi\ masses compared to the \hi\ mass of luminous galaxies \citep[based on the $\MHI - \mb$ relation; ][]{denes14mnras}; retaining them in the sample would thus significantly reduce the signal-to-noise ratio of the stacked \hii\ emission signal.} The cosmic volume occupied by our galaxy sample is $\approx 4.7 \times 10^4$ comoving Mpc$^3$.

Next, the presence of an active galactic nucleus (AGN) in a galaxy may affect its \hi\ properties. Outflows driven by AGNs may reduce the \hi\ mass of the AGN host galaxies \cite[e.g.][]{rupke11apjl,morganti16aa}. Further, the \hii\ emission spectra of the AGN hosts may also have associated \hii\  absorption features which could reduce the stacked \hii\ emission signal \citep[e.g.][]{morganti18aapr,aditya18cjf}. We hence excluded 84 galaxies identified as AGN hosts from the stacking sample, based on either their optical spectra \citep{cooper12mnras}, or a detection in our radio continuum image with $L_{1.4\:\mathrm{GHz}}\geq 2\times 10^{23} \:\mathrm{W\:Hz^{-1}}$ \citep[as 1.4~GHz radio luminosities above this threshold predominantly arise from radio-loud AGNs; ][]{condon02aj, smolcic08apj}. Finally, we restricted our sample to blue star-forming galaxies, by excluding 101 ``red'' galaxies, 
with $(\textrm{U} - \textrm{B}) + 0.032\:(M_{\textrm B}+21.62) - 1.035 > 0$ \citep{willmer06apj, coil08apj}.
Our final sample contains 623 blue star-forming galaxies at $z = 0.20 - 0.42$. 

The stellar masses of the 623 galaxies were estimated from their $\rm U - B$ colours, absolute B-band magnitudes ($\mb $), and redshifts ($z$), using the relation
\begin{equation}
{\rm log(M_*/M_{\odot})} = 1.39 \, {\rm (U-B)} - 0.44 \, {\rm M_B} - 2.10\, z + 0.84,
\label{eqn_lsm}
\end{equation} 
derived following \citet{weiner09apj}. This relation was calibrated for 127~blue star-forming galaxies (excluding AGN hosts) that are present in both the DEEP2/DEEP3 catalogues and the CANDELS catalogue \citep{stefanon17apjs}; the CANDELS stellar mass estimates were used as reference values in the calibration \citep[see][for details]{berainprep}. We note that, for the 127 galaxies that were used to calibrate the relation, the RMS scatter between the stellar masses estimated using this relation and the CANDELS stellar mass estimates is 0.16~dex.
Star-formation rate (SFR) estimates based on H$_{\alpha}$ and H$_{\beta}$ line luminosities for 309 of the 623 galaxies of our sample were provided by Benjamin J. Weiner, based on DEEP2 spectroscopy\footnote{Note that these estimates were obtained following the prescriptions of  \citet{hopkins03apj}, and do not include a correction for dust extinction.}. The above SFR estimates are entirely consistent with the (redshift-dependent) main-sequence relation between stellar mass and SFR given by \citet{whitaker12apj},
\begin{equation}
\textrm {log\:(SFR)} = \alpha (\textrm {log}\:M_* - 10.5) + \beta
\label{eqn_main-sequence}
\end{equation} 
where $\alpha$ and $\beta$ vary with redshift as
\begin{equation*}
\alpha (z) = 0.70 - 0.13 z \;\;{\rm , and}\;\; \beta(z) = 0.38 + 1.14z - 0.19z^2.
\end{equation*}
We hence used the above main-sequence relation to estimate the SFRs of all 623 galaxies in our sample from their stellar mass estimates; the error on the SFRs is $\approx 0.25$~dex, the scatter in the above main-sequence relation for blue star-forming galaxies \citep{whitaker12apj}. We emphasize that these SFRs are estimated from the stellar mass estimates using the main-sequence relation, and are not independent estimates.

\subsection{\hii\ subcubes of the sample galaxies} \label{subsec:subcubes}

For each galaxy in our sample, a subcube centred at its sky position and redshifted \hii\ line frequency was extracted from each of the four spectral data cubes (corresponding to the four observing cycles). The four subcubes obtained for each galaxy were treated independently in our analysis to ensure that any systematic errors in one of the observing cycles would not affect the subcubes from other cycles \citep[e.g.][]{chowdhury22survey}. Using the luminosity distance, $d_L(z)$, for each galaxy, each subcube was converted from observed flux density ($F_{\nu}$, in Jy) to spectral luminosity density ($L_{{\rm H}{\textsc i}}$, in Jy~Mpc$^2$), via the relation $L_{{\rm H}{\textsc i}} = {4 \pi \ F_{\nu} \ d_L^2(z)}/(1+z)$. 

Next, the optimal spatial resolution for the stacking analysis was chosen such that coarser spatial resolutions yield a stacked \hii\ emission signal consistent with that obtained at the optimal resolution. We found that the optimum spatial resolution for our sample is 30~kpc \citep[see][for detailed discussions]{bera21phd,berainprep}. Each subcube was convolved with a Gaussian beam to a (resultant) spatial resolution of 30~kpc at the redshift of the target galaxy, with appropriate normalization to ensure flux conservation \citep[i.e. by ensuring that the peak of the resultant point spread function is unity, after convolving with the same kernel;][]{chowdhury22survey}. Each subcube was then interpolated, spatially and spectrally, to a pixel size of 5~kpc (at the galaxy redshift) and a velocity resolution of 30~\kms. A second-order polynomial was then fitted to, and subtracted from, the spectrum at each spatial pixel of each subcube, excluding the central $\pm 200$~\kms\ from the fit. The final subcubes cover a spatial range of $500 \ {\rm kpc} \  \times 500 \ {\rm kpc}$, and a velocity range of $\pm 1000$~\kms, at the redshift of each galaxy. 

A set of statistical tests were next performed on each subcube to test for the presence of systematic effects (due to, e.g., RFI, deconvolution errors, etc.). Subcubes that showed evidence of such systematics were excluded from the sample. 

Finally, 14 galaxies with ``close neighbours'' were identified and excluded from the sample to reduce the possible effects of source confusion on our results. A ``close neighbour'' was defined as any galaxy with $\mb \leq -16$ within $30$~kpc and  $\pm$ 300 \kms~of the target galaxy. Our final sample for the stacking analysis consisted of 464 unique galaxies with 1665 subcubes, at a median redshift of $ z_{\rm med} = 0.35$.

\subsection{The \hii\ stacking procedure} \label{subsec:method}

\begin{figure*}[t]
\begin{center}
\includegraphics[scale=0.58, trim={0.4cm 0cm 0.3cm 0.2cm}, clip]{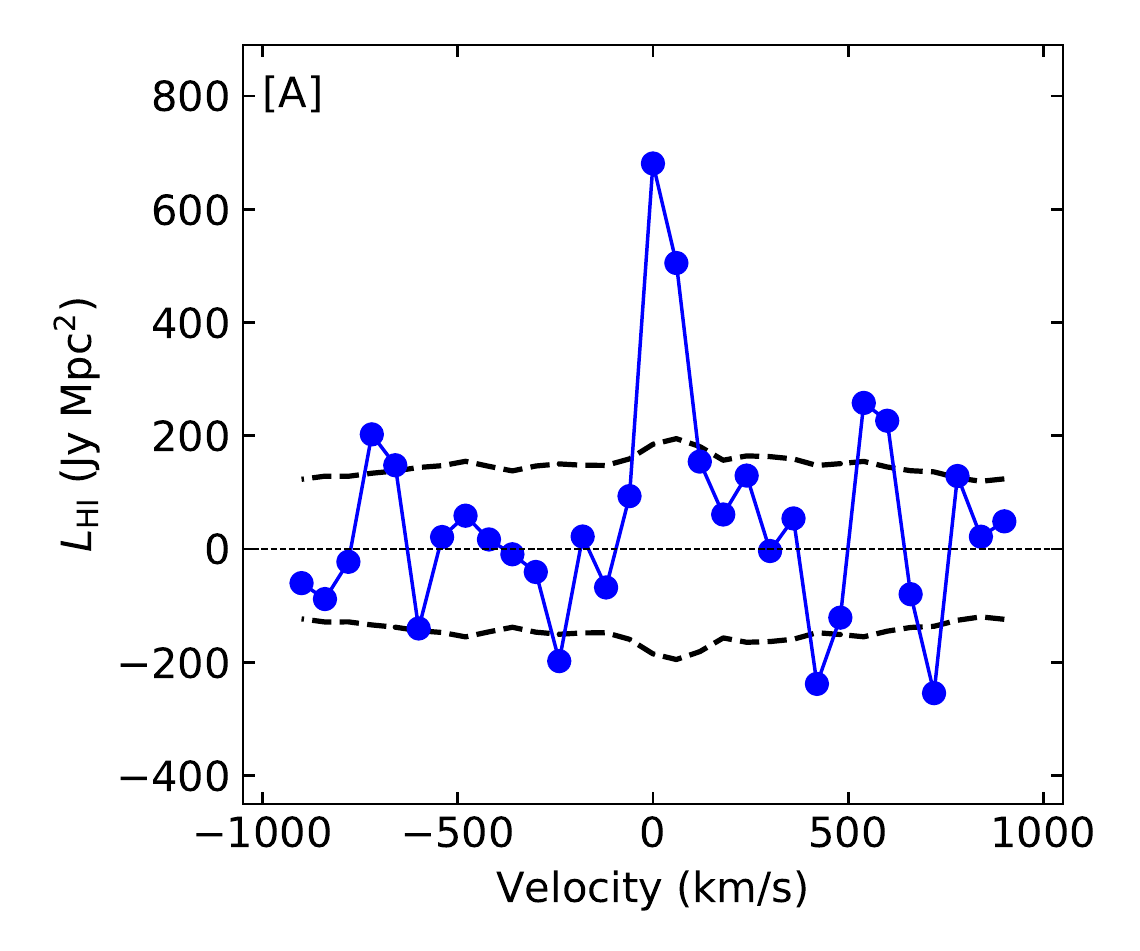}
\includegraphics[scale=0.58, trim={1.4cm 0cm 0.3cm 0.2cm}, clip]{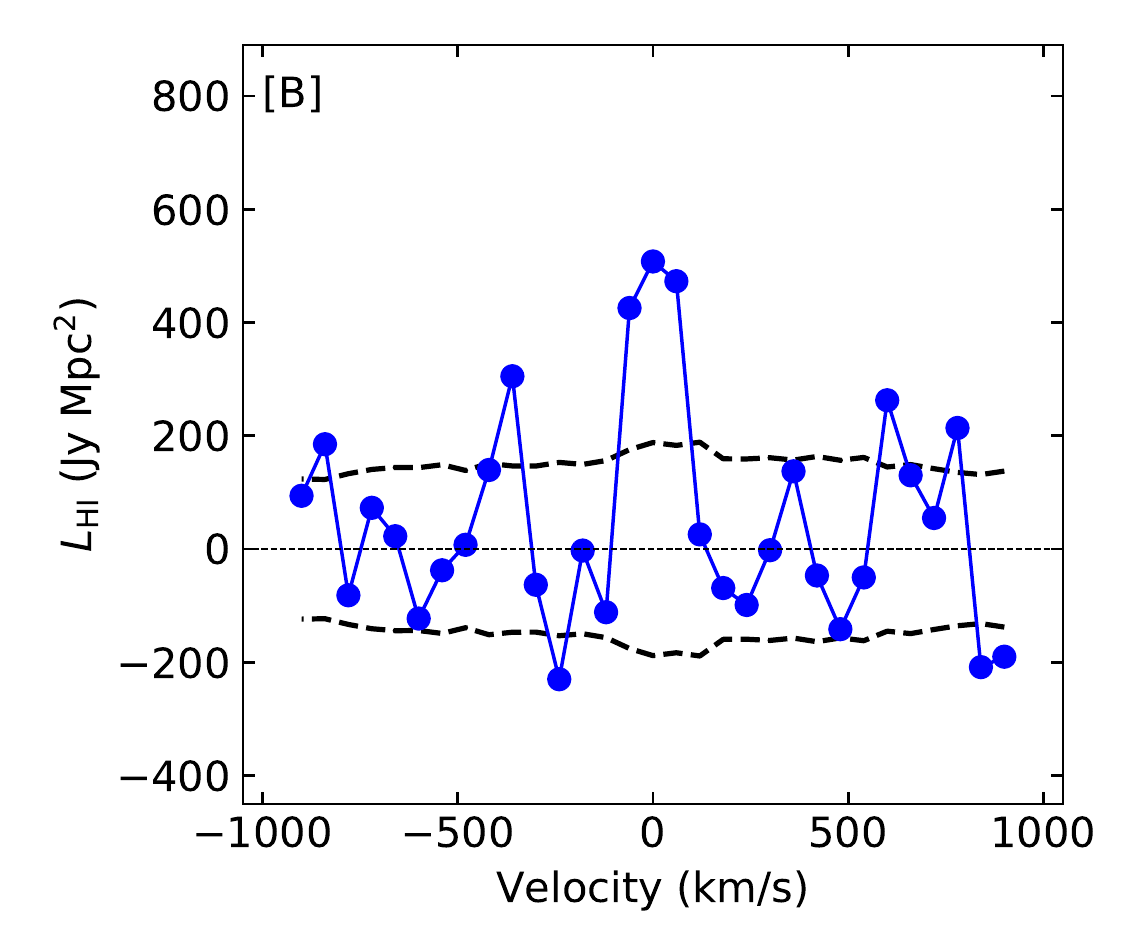}
\includegraphics[scale=0.58, trim={1.4cm 0cm 0.3cm 0.2cm}, clip]{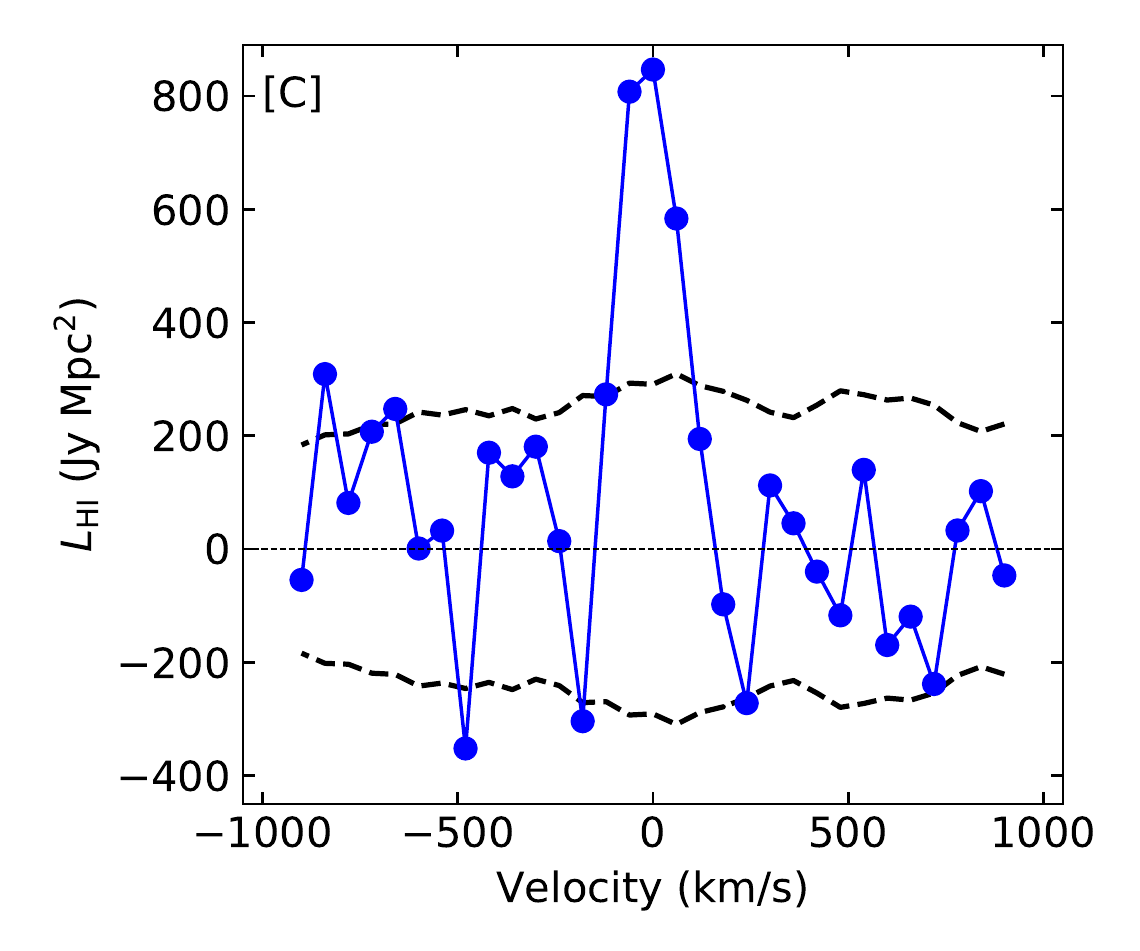}
\caption{The stacked \hii\ emission spectra (in units of luminosity density) for blue galaxies in three (logarithmic) stellar mass bins, [A]~$8.00 \leq \rm log\: (M_*/\msun ) < 8.70$, [B]~$8.70 \leq \rm log\: (M_*/\msun ) < 9.45$, and [C]~$9.45 \leq \rm log\: (M_*/\msun ) \leq 10.40$. The dashed black curves in each panel show the RMS noise in the corresponding velocity planes of the stacked spectral cubes. See text for discussion.}
\label{fig:spectra}
\end{center}
\end{figure*}

To determine the $\MHI - M_*$ relation at $z \approx 0.35$, we first divided the full galaxy sample into three independent (logarithmic) stellar mass bins. The number of stellar mass bins and the bin widths were selected to ensure that the average \hii\ emission signal is detected at $> 4\sigma$ significance in each bin. The stacking of the \hii\ emission signals was carried out independently for each stellar mass bin, by stacking the subcubes of all galaxies in the bin, plane-by-plane, with equal weights for all subcubes; this yielded a stacked spectral cube for each bin. A second-order polynomial was then fitted to, and subtracted out from, the spectrum at each spatial pixel of the three stacked cubes, again excluding the central $\pm 200$~\kms. The three residual stacked cubes were then Hanning-smoothed to, and resampled at, a velocity resolution of 60~\kms. For each stellar-mass bin, the final stacked spectrum was extracted at the central spatial pixel of the corresponding stacked cube.

The stacked \hii\ spectra for the three stellar-mass bins are shown in Figure~\ref{fig:spectra}; the stacked \hii\ emission signal is detected at $> 4\sigma$ significance in all three bins. The average \hii\ line luminosity for each bin was determined by integrating the stacked \hii\ spectrum over all contiguous central spectral channels with $\geq 1.5\sigma$ significance. For each bin, the RMS noise values measured from each plane of the spectral channels that were identified as ``line'' channels in the corresponding stacked \hii\ cube were combined to determine the measurement error, to estimate the detection significance of the average \hii\ signal. The average \hi\ mass of the galaxies in each stellar-mass bin was estimated from the corresponding stacked \hii\ line luminosity via the relation 
\begin{equation}
\frac{\MHI}{\rm M_\odot}  = 1.86 \times 10^{4} \times \frac{\int L_{{\rm H}{\textsc i}}\: dv}{\rm Jy\:Mpc^2\:km\:s^{-1}}.
\label{eqn_mh}
\end{equation}
Jackknife re-sampling was used to estimate the uncertainties on the average \hi\ masses. The jackknife errors are larger than the measurement errors, and include contributions from both sample variance and any underlying systematic effects. The average \hi\ masses and associated uncertainties are listed in Table~\ref{table:mhi-lsm}, along with the stellar mass ranges and the velocity-integrated \hii\ line luminosities for each bin. In passing, we note that slight changes in the stellar-mass ranges of the three bins were found to have no significant effect on the inferred scaling relation.

\begin{table*}
\begin{center}
\caption{\textbf{The average \hi~mass of blue galaxies in different (logarithmic) stellar mass  bins.} }
\begin{tabular}{|c|c|c|c|c|}
\hline
$\rm log\: (M_*/\msun ) $ range & Number of galaxies & Median $\rm M_*  $ & $ \int L_{{\rm H}{\textsc i}}\: dv$ & $\langle \MHI \rangle$ \\
 &     & $10^9 \; \msun$  & $10^5\: {\rm Jy\ Mpc^2\ km\ s^{-1}}$ & $10^9\:{\rm M}_\odot$ \\
\hline
$[8.00 - 8.70]$ & 198 & $0.26$ & $0.71 \pm 0.16$ & $1.33 \pm 0.32$ \\
$[8.70 - 9.45]$ & 174 & $1.00$ & $0.84 \pm 0.19$ & $1.57 \pm 0.41$ \\
$[9.45 - 10.40]$ & 67  & $6.75$ & $1.34 \pm 0.31$ & $2.50 \pm 0.69$ \\
\hline
\end{tabular}
\label{table:mhi-lsm}
\vskip 0.1in
For each stellar-mass bin, the columns are (1)~the stellar mass range, (2)~the number of galaxies, (3)~the median stellar mass, (4)~the velocity-integrated stacked \hii\ line luminosity, and (5)~the average \hi\ mass. The quoted errors on the velocity-integrated \hii\ line luminosities are  measurement errors, while those on the \hi\ mass are jackknife errors.
\end{center}
\end{table*}

\section{The $\rm \MHI- M_*$ scaling relation at $\mathit{z \approx 0.35}$} 
\label{sec:scaling}

\subsection{The ``mean'' $\mh - M_*$ scaling relation from \hii\ stacking and measurements of $\rm log\langle \mh \rangle$}

The $\mh - M_* $ scaling relation in the local Universe is well described by a linear relation between the logarithm of the \hi\ mass, $\rm log[\mh]$, and the logarithm of the stellar mass, $\rm log[M_*]$ \citep[e.g.][]{catinella10mnras, catinella18mnras, parkash18apj}. Since measurements of the \hi\ masses of individual galaxies are available in the nearby Universe, the best-fitting scaling relation is found using individual measurements of $\rm log[\mh]$; this is the ``standard'' approach. Conversely, our stacking analysis at $z \sim 0.35$ yields the average \hi\ mass of blue galaxies in different stellar mass bins, and does not yield measurements of the \hi\ masses of individual galaxies. The scaling relations from such stacking analyses are hence obtained by fitting to measurements of $\rm log \langle \mh \rangle$ in multiple stellar-mass bins \citep[e.g.][]{brown15mnras,chowdhury22scaling}. Thus, care must be taken when comparing the scaling relations from \hii\ stacking to those obtained from individual $\mh$ measurements at $z \approx 0$. In the present subsection, we will self-consistently compare scaling relations obtained by fits to measurements of $\rm log \langle \mh \rangle$ at all redshifts; we will refer to this as the ``mean'' $\mh - M_*$ scaling relation. The next subsection will discuss how one might determine the standard $\rm \mh - M_*$ scaling relation (referred to as the ``median'' $\mh - M_*$ relation) via stacking analyses.

We followed the optimization approach of \citet{chowdhury22scaling} to determine the mean $\mh - M_*$ scaling relation for blue galaxies at $z \approx 0.35$ from our measurements of the average \hi\ masses in three stellar mass bins. We initially assume that the mean $\mh - M_*$ relation can be written as
\begin{equation}
\rm log[\mh/\msun] = m_{9} + b\,[log (M_*/\msun) - 9.0],
\label{eqn_mhms}
\end{equation}
where $b$ is the slope of the relation and $m_9$ is a normalization constant.\footnote{Note that we use this form of the scaling relation to minimize the covariance between the two fitted parameters of the $\mh - M_*$ relation. Using the standard form $\rm log[\mh/\msun] = \alpha + \beta*log[M_*/\msun]$ \citep[e.g.][]{catinella18mnras,sinigaglia22apjl} gives a high covariance between $\alpha$ and $\beta$ when fitting to a given narrow range of stellar masses.} For given values of $b$ and $m_9$, the \hi\ mass of a galaxy, $\mh([b, m_9]; M_*)$, can be obtained from its (known) stellar mass $\rm M_*$ using Equation~\ref{eqn_mhms}. The average \hi\ mass of galaxies in different stellar mass bins can be calculated from these individual \hi\ masses. We denote the average \hi\ mass of the $i$'th stellar mass bin, corresponding to specific values of $b$ and $m_9$, by $\left\langle {\rm M}_{\textrm{\hi}} \right\rangle _i^{b,m_9}$. We find the best-fitting values of $b$ and $m_9$ by minimizing the quantity 
\begin{equation}
\chi^2 = \sum\limits_{i} \frac{[\left\langle {\rm M}_{\textrm{\hi}} \right\rangle _i^{b,m_9} - \left\langle {\rm M}_{\textrm{\hi}} \right\rangle _i]^2}{[\Delta \left\langle {\rm M}_{\textrm{\hi}} \right\rangle _i]^2} 
\label{eqn_chi2}
\end{equation}
where $\langle \mh \rangle_i$ is the measured average \hi\ mass of galaxies in the $i$'th bin and $\Delta \langle \mh \rangle_i$ is the jackknife error on this quantity. Note that, in this method, any intrinsic scatter of the \hi\ masses around the $\rm \mh - M_*$ scaling relation has been ignored (or, equivalently, assumed to be zero). 

\begin{figure*}[t]
\centering
\includegraphics[scale=0.66, trim={0.5cm 0.5cm 0 0}, clip]{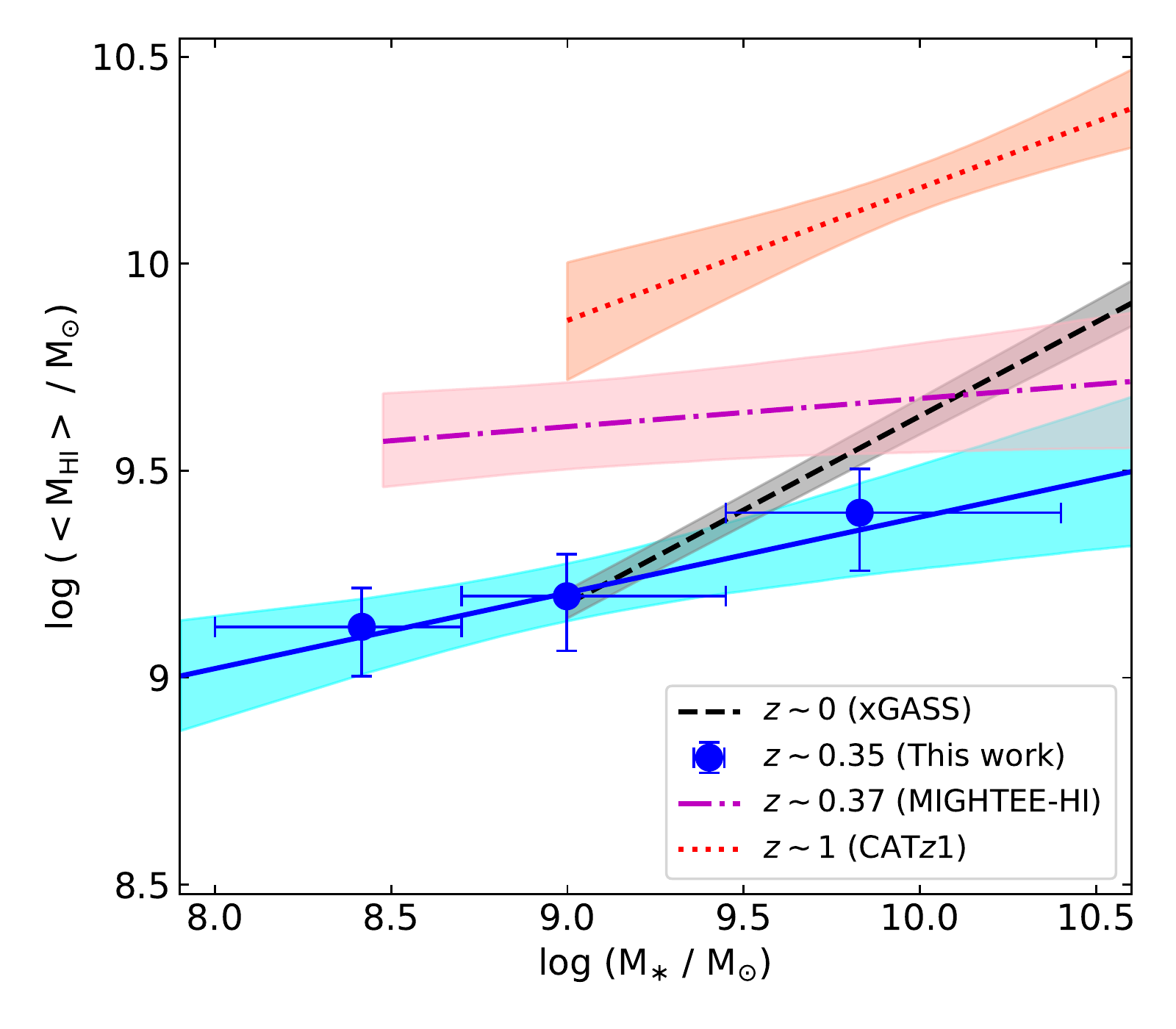}
\caption{\textbf{The} $\mathbf{\mh - M_*}$ \textbf{scaling relation, based on measurements of} $\mathbf{ \langle \mh \rangle}$ \textbf{in blue star-forming galaxies:} The blue filled circles show our \hii\ stacking measurements of $\rm log\, [\langle \mh \rangle/\msun]$ in three (logarithmic) stellar-mass bins for blue galaxies at $z \sim 0.35$ in the EGS. The solid blue line is the best-fit mean $\mh - M_*$ scaling relation, with the light blue shaded region showing the 68\% confidence interval around the fit. The figure also shows the mean $\mh - M_*$ scaling relations for blue star-forming galaxies at (1)~$z \approx 0$ from the xGASS sample \citep[dashed line;][]{catinella18mnras}, (2)~$z \approx 1$ from the GMRT-CAT$z$1 survey \citep[dotted line;][]{chowdhury22survey}, and (3)~at $z \approx 0.37$ from the MIGHTEE survey \citep[dash-dotted line;][]{sinigaglia22apjl}. Note that all  scaling relations in the panel have been obtained by fitting a relation to the dependence of $\rm log\; \langle \mh \rangle$ on $\rm M_*$, and not to $\rm \langle log\; \mh \rangle$, as is usually done at $z \approx 0$ \citep[e.g.][]{catinella18mnras}. 
}
\label{fig:mhi-mstar}
\end{figure*}

For blue galaxies at $z \sim 0.35$, we find that the best-fit mean $\rm \mh - M_*$ scaling relation is given by
\begin{equation}
\rm log\, [ \langle \mh \rangle /\msun] = (9.205 \pm 0.069) + (0.183 \pm 0.104)\:[log (M_*/M_{\odot}) - 9.0] \, ,
\label{eqn:mhi-mstar}
\end{equation}
which is shown as the blue solid line in Figure \ref{fig:mhi-mstar}. 

In the local Universe, \citet{catinella18mnras} obtained the standard $\mh - M_*$ scaling relation for the xGASS galaxy sample, using $\mh$ measurements in individual galaxies. To carry out a direct comparison to our scaling relation in Equation~\ref{eqn:mhi-mstar}, we selected 592 blue galaxies from the xGASS sample (satisfying ${\rm NUV} - r < 4$) and divided them in (logarithmic) stellar mass bins of width 0.2~dex.\footnote{We note that the xGASS sample only contains galaxies with $\rm{log \: (M_*/M_{\odot})} > 9.0$ \citep{catinella18mnras}.} We then calculated the mean \hi\ mass, $\langle \mh \rangle$, for each stellar-mass bin, and estimated the associated uncertainty by jackknife resampling. Using these average \hi\ masses and following the approach described above, we find that the best-fit mean $\mh - M_*$ relation for blue xGASS galaxies is
\begin{equation}
{{\rm log\, [ \langle \mh \rangle /\msun]} = (9.178 \pm 0.035) + (0.454 \pm 0.027)\:[{\rm log (M_*/M_{\odot}) - 9.0}]}.
\label{eqn:xgass}
\end{equation} 
This relation is shown as the dashed line in Figure \ref{fig:mhi-mstar}. 

While comparing our scaling relation at $z \approx 0.35$ with that obtained from the xGASS sample at $z \approx 0$, it is important to emphasize that the stellar mass distributions of the two samples are different (as can be clearly seen in the stellar mass ranges of the two relations in Fig.~\ref{fig:mhi-mstar}). A comparison between the scaling relations requires us to assume that the xGASS relation may be extrapolated to lower stellar masses, $\approx 10^8 \ \msun$, and that the EGS relation is applicable at high stellar masses, $\approx 10^{11} \ \msun$.

With this assumption, we find that the normalization constants of the scaling relations at $z \sim 0.35$ and $z \sim 0$ are consistent within the errors, but the slopes are different at $\approx 2.5 \sigma$ significance, with the relation at $z \sim 0.35$ being flatter than the local relation. This result indicates that massive star-forming galaxies with $\rm M_* \gtrsim 10^{10} \, \msun$ are relatively \hi-poor compared to their local counterparts. Conversely, low-mass star-forming galaxies at $z \approx 0.35$, with $\rm M_* < 10^{8.5} \, \msun$, appear to be \hi-rich compared to similar galaxies at $z \approx 0$.

The dotted line in Fig.~\ref{fig:mhi-mstar} shows the mean $\mh - M_*$ scaling relation at $z \approx 1$, obtained from \hii\ stacking with the GMRT-CAT$z$1 survey \citep{chowdhury22survey,chowdhury22scaling}. The slope of the scaling relation at $z \approx 1$ is formally intermediate between the values at $z \approx 0.35$ (obtained here) and $z \approx 0$, albeit consistent with both, within the errors (we again assume that each scaling relation may be extrapolated to the stellar mass range of the other relations). However, the normalization of the relation is significantly higher at $z \approx 1$, by $\approx 0.6$~dex at $\rm M_* \approx 10^{10} \, \msun$.
This indicates that star-forming galaxies at $z \approx 1$ are $\approx 3.5$ times more gas-rich than galaxies of similar stellar masses at $z \approx 0.35$ and $z \approx 0$ \citep{chowdhury22scaling}.

Recently, \citet{sinigaglia22apjl} used a similar \hii\ stacking approach with MeerKAT observations of the COSMOS field to measure the mean $\mh - M_*$ scaling relation for star-forming galaxies at $z \approx 0.37$. We used their  average \hi\ mass measurements in different stellar mass bins (kindly provided by F.~Sinigaglia) to obtain the best-fit scaling relation of the form given in Equation~\ref{eqn_mhms}. This yields $\rm log[\langle \mh \rangle/\msun] = (9.61 \pm 0.11) + (0.068 \pm 0.078)\; [log(M_*/\msun)-9.0]$ at $z \approx 0.37$. Figure~\ref{fig:mhi-mstar} compares our $\rm \mh - M_*$ relation (blue solid line) with that of \citet{sinigaglia22apjl} (dash-dotted line). The scaling relation of \citet{sinigaglia22apjl} is seen to lie above our relation (by $\approx 0.4$~dex at $\rm M_* \approx 10^9 \, \msun$), with an $\approx 3 \sigma$ difference in the normalization of the relation. Finally, while the slope of the COSMOS scaling relation in Fig.~\ref{fig:mhi-mstar} appears flatter than that of the EGS scaling relation, the two slopes are formally consistent within the errors.

Our present measurement has a normalization similar to that of the relation at $z \approx 0$ (from the xGASS sample), while the relation of \citet{sinigaglia22apjl} lies above the local relation. The two scaling relations have different implications for the evolution of gas in galaxies from $z \approx 1$ to the present epoch. We emphasize that the scaling relation obtained here, from the EGS, is based on a relatively small cosmic volume ($\approx 4.7 \times 10^4$ comoving Mpc$^3$) and a small number of galaxies ($\approx 500$ objects), implying that cosmic variance could affect our results. However, the detection significance in each stellar-mass bin is relatively high, with $>4\sigma$ significance. Conversely, the result of \citet{sinigaglia22apjl} is based on a far larger sample ($\approx 9000$~galaxies) and a larger cosmic volume \citep[$\approx 8.5\times 10^5$ comoving Mpc$^3$, albeit still significantly affected by cosmic variance; e.g.][]{driver10mnras}. However, there are clear systematic effects in the stacked \hii\ spectra of \citet{sinigaglia22apjl}, including oscillations in both the full stacked spectrum and the reference spectrum of their Fig.~2, an absorption feature stronger than the emission feature in the top middle spectrum of their Fig.~3, etc, and the detection significance of the stacked \hii\ emission signal is low ($\approx 3\sigma$) in some bins. All of these make the results less reliable. In addition, \citet{sinigaglia22apjl} do not exclude objects identified as AGNs from their sample ($\approx 3.5$\% of their galaxies); this too could affect the slope of the $\mh - \Mstar$ relation, if AGNs predominantly arise in more massive galaxies. Overall, a combination of deeper and wider \hii\ stacking observations is critically needed to accurately determine the shape of the $\rm \mh - \Mstar$ relation at intermediate redshifts and to trace the evolution of \hi\ in galaxies from $z \approx 1$ to today.

\subsection{The ``median'' $\mh - M_*$ scaling relation from \hii\ stacking and estimates  of $\rm \langle log \mh \rangle$}

In the previous subsection, we presented the mean $\mh - M_*$ scaling relation obtained by fitting to the $\rm log \langle \mh \rangle$ values that are measured in \hii\ stacking studies. As noted earlier, in the local Universe, one usually determines the $\mh - M_*$ relation by fitting to measurements of the stellar mass and the \hi\ mass of individual galaxies; this is effectively the same as fitting to measurements of $\rm \langle log\, \mh \rangle$. For the blue galaxies of the xGASS sample, the best-fit $\mh - M_*$ relation obtained by this approach is \citep{catinella18mnras}
\begin{equation}
\rm log\, (\mh/\msun) = (8.934 \pm 0.036) + (0.516 \pm 0.030)\:[log\,(M_*/\msun) - 9.0]
\label{eqn:xgass2}
\end{equation} 
This scaling relation has a logarithmic scatter of $\sigma = 0.445$~dex \citep{catinella18mnras}.

A comparison between Equations~\ref{eqn:xgass}  and \ref{eqn:xgass2} shows that the slopes of the two $\mh - M_*$ relations at $z \approx 0$ are in excellent agreement, but that the former relation lies $\approx 0.25$~dex above the latter one. This is to be expected when there is an intrinsic scatter in the \hi\ masses in each stellar-mass bin: for a symmetric logarithmic scatter in the \hi\ masses, the relation in Equation~\ref{eqn:xgass} traces the logarithm of the average \hi\ mass in each stellar-mass bin, while the relation in Equation~\ref{eqn:xgass2} traces the logarithm of the median \hi\ mass in each bin \citep{bera22apjl}. We will hence refer to the standard $\mh - M_*$ scaling relation, obtained by fitting to measurements of the \hi\ mass and the stellar mass in individual galaxies, as the median $\mh - M_*$ relation. 

As noted by \citet{bera22apjl}, one needs to know the intrinsic scatter in the $\mh - M_*$ relation in order to determine the median $\mh - M_*$ relation from the mean $\mh - M_*$ relation. Unfortunately, this information on the intrinsic scatter in the $\mh - M_*$ relation is not presently available beyond the local Universe.

To determine the median $\mh - M_*$ relation at $z \approx 0.35$ from our measurements of the average \hi\ masses in the different stellar mass bins, we assume that the \hi\ masses of individual blue galaxies in each stellar-mass bin at $z \sim 0.35$ are lognormally distributed, with the same intrinsic scatter as that of blue xGASS galaxies at $z \approx 0$. In other words, we assume that the intrinsic scatter in the $\mh - M_*$ relation for blue galaxies at $z \approx 0.35$ is $\sigma=0.445$~dex, the same as that in the local Universe. \citet{bera22apjl} show that the ratio of the mean \hi\ mass to the median \hi\ mass for a sample of galaxies can be used to obtain an upper limit to the intrinsic lognormal scatter in the $\mh - M_*$ relation. For our 464 blue galaxies at $z \approx 0.35$, the mean \hi\ mass is $(1.57 \pm 0.26)\times 10^9 \, \msun$, while the median \hi\ mass is $(1.17 \pm 0.22)\times 10^9 \, \msun$. Following \citet{bera22apjl}, we then obtain the upper limit $\sigma_{\rm max} = 0.33 \pm 0.10$~dex, which is consistent with the assumed scatter of 0.445~dex. 

For a lognormal distribution of \hi\ masses, with an intrinsic scatter $\sigma$, the logarithm of the mean of the distribution and the mean of the logarithm of individual \hi\ masses are related by
\begin{equation}
\rm{log\, \langle \mh \rangle - \langle log\, \mh \rangle = \frac{ln\:10}{2}\sigma^2}
\label{eqn:mean-median}
\end{equation} 

The best-fit median $\mh - M_*$ scaling relation can hence be obtained by combining Equations~\ref{eqn:mhi-mstar} and \ref{eqn:mean-median}; this gives
\begin{equation}
\rm log\, (\mh/\msun) = (8.977 \pm 0.069) + (0.183 \pm 0.104)\, [log (M_*/\msun) - 9.0].
\label{eqn:mhi-mstar-med}
\end{equation}
In passing, we note that a very similar relation is obtained by explicitly carrying out a least-squares fit via Equation~\ref{eqn_chi2}, but this time including the intrinsic lognormal scatter of $0.445$~dex.

We emphasize that the derivation of the median $\mh - M_*$ relation in Equation~\ref{eqn:mhi-mstar-med} makes the critical assumptions that the intrinsic scatter of \hi\ masses in each stellar-mass bin follows a lognormal distribution, and that the intrinsic scatter is the same as that in the local Universe. The mean $\mh - M_*$ relation of Equation~\ref{eqn:mhi-mstar} does not make these assumptions, and is thus a more general relation from \hii\ stacking studies.

\section{The \hi\ depletion timescale} \label{sec:depletiontimescale}

\begin{figure}[t]
\centering
\includegraphics[scale=0.645, trim={0.6cm 0.5cm 0.5cm 0}, clip]{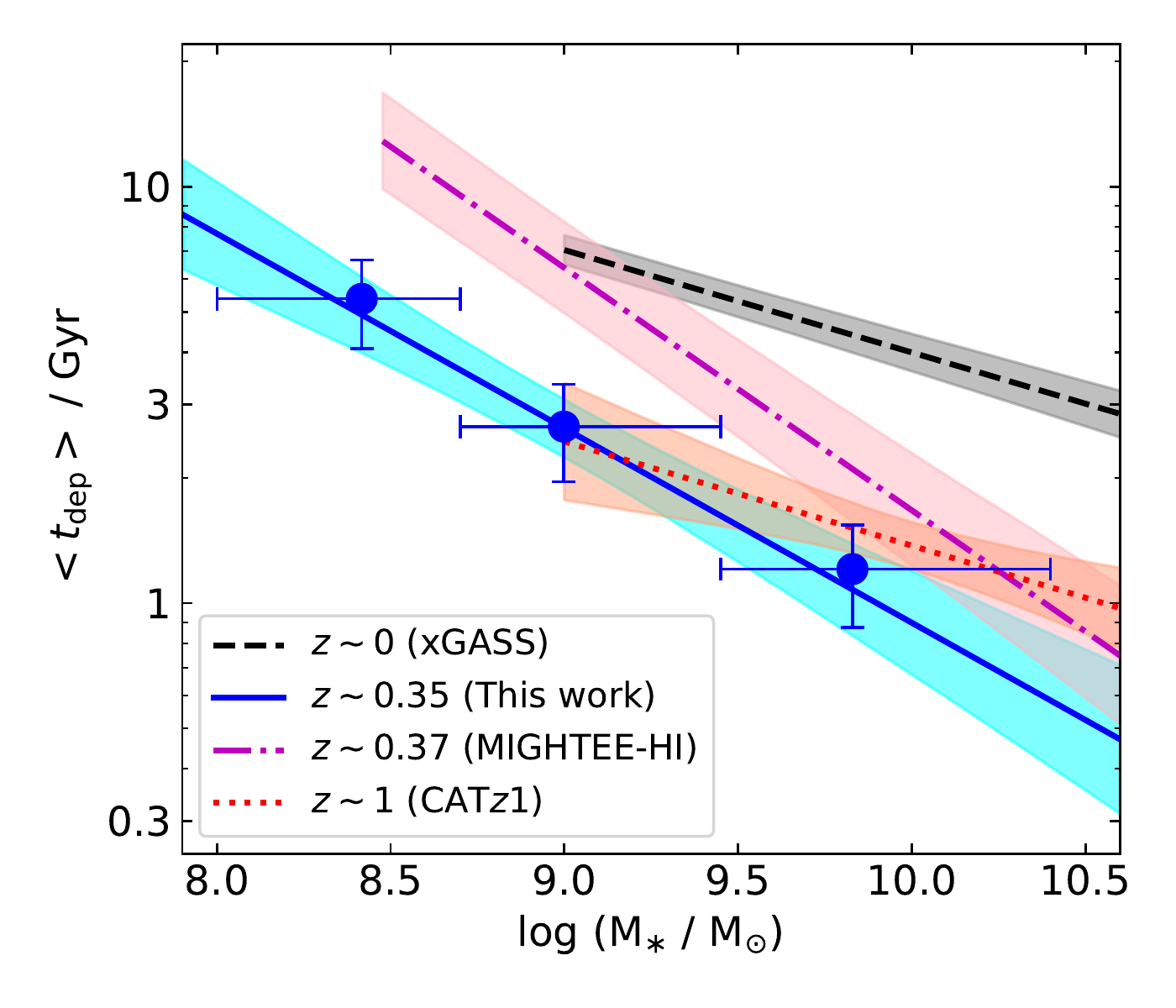}
\caption{\textbf{The characteristic \hi\ depletion timescale as a function of stellar mass at different redshifts:} The blue circles show the measurements of the characteristic \hi\ depletion times for blue galaxies at $z \approx 0.35$ for three stellar mass bins, while the blue solid line and shaded region show, respectively, the $t_{\rm dep} - \rm M_*$ scaling relation and the 68\% confidence interval for these galaxies. The figure also shows the characteristic \hi\ depletion timescales for blue star-forming galaxies at (1)~$z \approx 0$ from the xGASS sample \citep[dashed black line;][]{catinella18mnras}, (2)~$z \approx 1$ from the GMRT-CAT$z$1 survey \citep[dotted red line;][]{chowdhury22survey}, and (3)~$z \approx 0.37$ from the MIGHTEE-HI survey \citep[dash-dotted magenta line;][]{sinigaglia22apjl}. Note that all  scaling relations in the panel have been obtained uniformly by combining the $\rm log\; \langle \mh \rangle - log\; M_*$ relation with the (redshift-dependent) main-sequence relation from \citet{whitaker12apj}.}
\label{fig:tdep-ms}
\end{figure}

The \hi\ depletion timescale ($t_{\rm dep} \equiv \mh/{\rm SFR}$) of a galaxy gives the timescale over which its \hi\ mass would be completely consumed at its current SFR (with an intermediate conversion to molecular hydrogen). $t_{\rm dep}$ is thus  a measure of how long a galaxy can sustain its star-formation activity without replenishment of its \hi\ reservoir. Its inverse is a measure of the current star-formation efficiency of a galaxy, relative to its \hi\ content. In the local Universe, the \hi\ depletion timescale has been found to show only a weak dependence on the stellar mass \citep[e.g.][]{schiminovich10mnras,saintonge17apjs}.

As noted earlier, \hii\ stacking studies yield direct measurements of the average \hi\ mass of galaxy samples. One can combine these average \hi\ masses with the average SFRs of the same samples to obtain the {\it characteristic} \hi\ depletion timescale, $\langle t_{\rm dep} \rangle \equiv \rm \langle \mh \rangle / \langle SFR \rangle$, of the sample. We emphasize that the characteristic \hi\ depletion timescale is not the same as the average \hi\ depletion timescale, $\langle \mh / \rm SFR \rangle$ \citep[e.g.][]{chowdhury22scaling}.

The dependence of the characteristic \hi\ depletion timescale on the stellar mass can be obtained by combining Equation~\ref{eqn:mhi-mstar} with the main-sequence relation of star-forming galaxies \citep[e.g.][]{brinchmann04mnras,noeske07apj,whitaker12apj}. At $z = 0.35$, the median redshift of our galaxy sample, the main sequence is well described by the relation \citep[][]{whitaker12apj} 
\begin{equation}
\rm {log\:(SFR / M_{\odot}\: yr^{-1})} = 0.65\: [\textrm {log}\:(M_*/M_{\odot}) - 10.5] + 0.76 \;.
\label{eqn:ms35}
\end{equation} 
Dividing Equation~\ref{eqn:mhi-mstar} by Equation~\ref{eqn:ms35}, we find that the characteristic \hi\ depletion timescale for blue galaxies at $z \sim 0.35$ is given by
\begin{equation}
{{\rm log}\: [ \langle t_{\rm dep} \rangle / \rm Gyr]} = (0.420 \pm 0.069) - (0.467 \pm 0.104)\:[{\rm log (M_*/M_{\odot}) - 9.0}] \; .
\label{eqn:tdepmsz35}
\end{equation}
Note that the uncertainties in this relation do not include the uncertainties in the slope and the normalization of the main-sequence relation (Equation~\ref{eqn:ms35}).

Following a similar approach, we estimated the characteristic \hi\ depletion timescales for blue star-forming galaxies at $z \approx 0$ and $z \approx 1$, by combining the corresponding  $\rm \mh - M_*$ scaling relation with the (redshift-dependent) main-sequence relation  \citep{whitaker12apj}. We obtain, for blue galaxies in the xGASS sample at  $z \approx 0$ \citep[][]{catinella18mnras}
\begin{equation}
{{\rm log}\: [ \langle t_{\rm dep} \rangle / \rm Gyr]} = (0.848 \pm 0.035) - (0.246 \pm 0.027)\:[{\rm log (M_*/M_{\odot}) - 9.0}] \; ,
\label{eqn:tdepmsz0}
\end{equation}
while, for blue galaxies at $z \approx 1$ from the GMRT-CAT$z$1 survey \citep{chowdhury22survey},\footnote{We note that \citet{chowdhury22scaling} use a second-order form of the main-sequence relation, from \citet{whitaker14apj}, to obtain the $\langle t_{\rm dep}\rangle - \rm M_*$ relation at $z \approx 1$.}
\begin{equation}
{{\rm log}\: [ \langle t_{\rm dep}\rangle / \rm Gyr]} = (0.138 \pm 0.056) - (0.25 \pm 0.13)\:[{\rm log (M_*/M_{\odot}) - 10.0}] \; .
\label{eqn_tdepmsz1}
\end{equation}

Figure~\ref{fig:tdep-ms} compares the $\langle t_{\rm dep} \rangle - \rm M_*$ relation obtained from our observations of the EGS (solid blue line) with the relations obtained at $z \approx 1$ (dotted line) and $z \approx 0$ (dashed line). The shaded regions indicate the 68\% confidence interval around each relation.  We find that the characteristic \hi\ depletion timescales of star-forming galaxies (with $\rm M_* > 10^{9} \; M_{\odot}$) at $z \sim 0.35$ are lower by a factor of $\approx 3-5$ compared to those of galaxies with similar stellar masses at the present epoch. However, the characteristic \hi\ depletion timescales of blue galaxies at $z \sim 0.35$ in the EGS are comparable to those of blue galaxies of similar stellar masses at $z \sim 1$ \citep[][]{chowdhury22survey}. Our current results thus suggest that the star-formation efficiency (relative to \hi) of blue star-forming galaxies did not evolve significantly over the $\approx 4$~Gyr from $z \approx 1$ to $z \approx 0.35$, but declined steeply over the next $\approx 4$~Gyr, from $z \approx 0.35$ to $z \approx 0$. 

Figure~\ref{fig:tdep-ms} shows that star-forming galaxies at $z \approx 0.35$, roughly 4~Gyr ago, with $\Mstar \gtrsim 10^9\ \msun$ have characteristic \hi\ depletion timescales lower than $\approx 3$~Gyr. Such galaxies would entirely consume their \hi\ reservoirs much before the present epoch. This would result in  quenching of their star-formation activity, unless they accrete significant amounts of gas from the circumgalactic medium (CGM) over the past 4~Gyr, to replenish their \hi\ reservoirs. This is a very interesting result, although we again emphasize that it is based on observations of a relatively small cosmic volume ($\approx 4.7 \times 10^4$ comoving Mpc$^3$), and hence may be affected by cosmic variance.

Figure~\ref{fig:tdep-ms} also shows the characteristic \hi\ depletion timescale at $z \approx 0.37$ (dash-dotted line) obtained using the $\rm \mh - M_*$ scaling relation from the MIGHTEE-HI survey \citep{sinigaglia22apjl}. This yields ${{\rm log}\: \left[ \langle t_{\rm dep} \rangle / \rm Gyr \right]} = (0.81 \pm 0.11) - (0.582 \pm 0.078)\: \left[ {\rm log (M_*/M_{\odot}) - 9.0} \right]$. We find that the MIGHTEE-HI $\langle t_{\rm dep} \rangle - \rm M_*$ relation has a slope similar to ours, but a higher normalization, by a factor of $\approx 2.5$. The discrepancy is due to the difference in the two ``mean'' $\rm \mh - M_*$ relations, discussed in the previous section. As noted earlier, a combination of deep and wide \hii\ stacking observations is critical to resolve this issue.

\section{Summary} \label{sec:conclusion}

We have used a GMRT \hii\ survey of the Extended Groth Strip, and the technique of \hii\ stacking, to measure the scaling relation between the \hi\ mass and the stellar mass for blue star-forming galaxies at $z \approx 0.35$. We have also combined our estimate of the $\rm \mh - M_*$ relation with the main-sequence relation for star-forming galaxies at $z \approx 0.35$ to determine the dependence of the characteristic \hi\ depletion timescale of star-forming galaxies on their stellar mass. We find that massive star-forming galaxies at $z \approx 0.35$, with $\rm M_* \gtrsim 10^{10}\:M_{\odot}$, are relatively \hi-poor compared to local star-forming galaxies of similar stellar mass. However, the \hi-based star-formation efficiency of massive galaxies at $z \approx 0.35$ is higher by a factor of $\approx 5$ compared to that of their local counterparts.

The $\mh - \Mstar$ scaling relation obtained in our study of the EGS at $z \approx 0.35$ is different (at $\approx 3\sigma$ significance) from that obtained by \citet{sinigaglia22apjl} at $z \approx 0.37$ from a similar \hii\ stacking analysis applied to MIGHTEE-HI observations of the COSMOS field. Both studies stack the \hii\ signals from star-forming galaxies on the main sequence; the difference between the scaling relations is unlikely to arise from differences in the sample or the approaches to determine the $\Mstar$ estimates (although we note that, unlike the present study, \citet{sinigaglia22apjl} retain AGNs in their sample). 
The difference between the two results is likely to arise from cosmic variance (affecting both studies due to their relatively small cosmic volumes), and a combination of systematic errors and low significance measurements affecting the spectra of \citet{sinigaglia22apjl}.

While our results are based on a small cosmic volume and could thus be affected by cosmic variance, the short inferred characteristic \hi\ depletion timescales ($\lesssim 3$~Gyr) of star-forming galaxies at $z \approx 0.35$ with stellar masses $\gtrsim 10^9\ \Mstar$ imply that such galaxies must have acquired significant amounts of neutral gas through accretion from the CGM over the past four Gyr to prevent quenching of their star-formation activity.

\begin{acknowledgments}
We thank an anonymous referee for detailed comments on an earlier version of this paper. We thank the staff of the GMRT who have made these observations possible. The GMRT is run by the National Centre for Radio Astrophysics of the Tata Institute of Fundamental Research. We are grateful to Francesco Sinigaglia for providing us with the average \hi\ and average stellar masses in the different bins of their MIGHTEE-HI study, and to Benjamin Weiner for providing us with the H$\alpha$ and H$\beta$ line luminosities of the DEEP2 galaxies. AB and NK thank Aditya Chowdhury for many discussions on \hii\ stacking that have contributed to this paper. NK acknowledges support from the Department of Science and Technology via a Swarnajayanti Fellowship (DST/SJF/PSA-01/2012-13). AB, NK, $\&$ JNC also acknowledge the Department of Atomic Energy for funding support, under project 12-R\&D-TFR-5.02-0700. 
\end{acknowledgments}
\newpage
\bibliography{galaxyrefs}{}
\bibliographystyle{aasjournal}

\end{document}